\documentstyle[12pt]{article}
\newcommand{\e}{\; {\rm e} }
\newcommand{\sgn}{\; {\rm sgn} }
\newcommand{\tr}{\; {\rm tr} }
\newcommand{\be}{\begin{eqnarray} }
\newcommand{\ee}{\end{eqnarray} }

\begin{document}
\begin{center}
\large{\bf Dynamical Chern-Simons term generation at finite density}
\vskip 5mm
\large{A. N. Sissakian\\}
{\it Bogolubov Theoretical Laboratory,
Joint Institute for Nuclear Research,
Dubna, Moscow region 141980, Russia \\}
\vskip 5mm
\large{O. Yu. Shevchenko\footnote{e-mail: shevch@nusun.jinr.ru}
and S. B. Solganik\footnote{e-mail: solganik@thsun1.jinr.ru}\\}
{\it Laboratory of Nuclear Problems,
Joint Institute for Nuclear Research,
Dubna, Moscow region 141980, Russia\\}
\end{center}
\vskip 5mm

\begin{abstract}
The Chern-Simons topological term dynamical generation in the
effective action is obtained at arbitrary finite density.
By using the proper time method  and perturbation theory
it is shown  that $\mu^2 = m^2$ is the
crucial point for Chern--Simons.
So when $\mu^2 < m^2$
$\mu$--influence disappears and we get the usual Chern-Simons term.
On the other hand when $\mu^2 > m^2$ the Chern-Simons term
vanishes because of non--zero density of background fermions.
In particular for massless case parity anomaly is absent at any
finite density.
This result holds in any odd dimension as in abelian
so as in nonabelian cases.\\
\end{abstract}
\vskip 5mm

Since introducing the Chern-Simons (CS) topological term \cite{jackiw}
and by now  the great number of papers  devoted to it appeared.
Such interest is  explained by variety of
significant physical effects caused by CS secondary
characteristic class.
These are, for example, gauge particles mass appearance  in quantum
field theory, applications to condense matter physics  such as
the fractional quantum Hall effect and high $T_{c}$ superconductivity,
possibility of free of metric tensor theory construction
and so on.

It was shown [2-4] in a conventional zero density gauge
theory that the CS term is generated in the
Eulier--Heisenberg effective action
by quantum corrections.
The main goal of this paper is to explore the parity anomalous
CS term generation at finite density.
%%%%%%%%%%%%%%%%%%%%%%%%%%%%%%%%%%%%%%%%%%%%%%%%%%%%%%%%%%%%
In the excellent paper by Niemi \cite{ni} it was emphasized that the charge density
at $\mu \not = 0$ becomes nontopological object, i.e contains as topological
part so as nontopological one.
The charge density at $\mu \not = 0$ (nontopological, neither parity odd
nor parity even object)\footnote{For abbreviation,  speaking about parity
invariance properties of local objects, we will
keep in mind  symmetries of the corresponding action parts.}
in $QED_{3}$ at finite density
was calculated and exploited in \cite{tolpa}. It must be
emphasized that in \cite{tolpa} charge density contains
as well parity odd part corresponding to CS term
so as parity even part, which can't be covariantized
and don't contribute to the mass of the gauge field.
Here we are interested in effect of finite density influence
on covariant parity odd form in action leading to the
gauge field mass generation --- CS
topological term. Deep insight on this phenomena at small densities
was done in \cite{ni,ni1}.
The result for CS term coefficient in $QED_{3}$  is
$\left[ \sgn(m-\mu)+\sgn(m+\mu)\right]$
(see \cite{ni1}, formulas (10.19) ).
However, to get this result it was heuristicaly supposed
that at small densities index theorem could still be used and
only odd in energy part of spectral density is responsible for
parity nonconserving effect.
Because of this in \cite{ni1} it had been stressed
that the result holds only for small $\mu$. However,
as we'll see below this result holds for any values of
chemical potential.
Thus, to obtain trustful result at any values of $\mu$ one
have to use transparent and free of any restrictions on $\mu$
procedure,
which would allow to perform calculations
with arbitrary nonabelian background gauge fields.
%%%%%%%%%%%%%%%%%%%%%%%%%%%%%%%%%%%%%%%%%%%%%%%%%%%%%%%%%%%%

Since  the chemical potential term $\mu\bar\psi\gamma^{0}\psi$ is
odd under charge conjugation we can expect that it would
contribute to $P$ and $CP$ nonconserving quantity ---  CS term.
As we will see, this expectation is completely justified.

The zero density approach usually is a good quantum field  approximation
when the chemical potential is small as compared with
characteristic energy scale  of physical processes.
Nevertheless, for investigation of topological effects
it is not the case.
As we will see below, even a small density could lead to
principal effects.

Introduction of a chemical potential $\mu$ in a theory   corresponds
to the presence of a non\-va\-ni\-shing background charge density.
So, if $\mu >0$, then the number of particles exceeds  that of
antiparticles and vice versa. It must be emphasized that the
formal addition of a chemical potential looks like a simple gauge
transformation
with the gauge function $\mu t$. However, it doesn't only shift the time
component
of a vector potential but also gives corresponding prescription for
handling Green's function poles.
The correct introduction of a chemical potential redefines
the ground state (Fermi energy),
which leads to a new spinor propagator with the correct
$\epsilon$-prescription for poles.
So, for the free spinor propagator we have
(see, for example, \cite{shur,chod})
\begin{eqnarray}
G(p;\mu)=
\frac{\tilde{\not\! p}+m}
{(\tilde{p_{0}}+i\epsilon\sgn p_0 )^2-\vec{p}^2-m^2 },
\end{eqnarray}
where $\tilde{p}=(p_0 + \mu,\vec{p})$. Thus, when $\mu =0$ one at once
gets the usual $\epsilon$-prescription  because of the positivity of
$p_0\sgn p_0$.
In the presence of a  background Yang--Mills field we consequently
have for the Green function operator
\begin{eqnarray}
\hat{G}=(\gamma \tilde{\pi}-m)\frac{1}{(\gamma\tilde{\pi})^2-m^2 + i \epsilon
(p_{0}+\mu)\sgn (p_{0})} ,
\end{eqnarray}
where $\tilde{\pi_{\nu}}=\pi_{\nu}+\mu\delta_{\nu 0}$ ,
$\pi_{\nu}=p_{\nu}-gA_{\nu}(x)$.

Let's first consider a (2+1) dimensional  abelian case and choose
the background field in the form
$$ A^{\mu}=\frac{1}{2}x_{\nu}F^{\nu\mu} ,
\;\;\;\;\;\; F^{\nu\mu}={\rm Const}. $$
To obtain the  CS term in this case, it is necessary to consider
the background current
$$ \langle J^{\mu}\rangle =  \frac{\delta S_{eff}}{\delta A_{\mu}}  $$
rather than the effective action itself. This is because the CS
term formally vanishes for such the choice of
$A^{\mu}$  but its variation with respect to $A^{\mu}$
produces a nonvanishing current.
So, consider
\begin{equation}
\langle J^{\mu}\rangle= -i g \tr\left[\gamma^{\mu}G(x,x^{'})
\right]_{x\rightarrow x^{'}}
\end{equation}
where
\begin{equation}
G(x,x^{'})=\exp\left(-ig\int_{x^{'}}^{x}d\zeta_{\mu}A^{\mu}(\zeta)\right)
\langle x | \hat G | x^{'}\rangle .
\end{equation}
Let's rewrite  Green function (2) in a more appropriate form
\begin{eqnarray}
\hat{G}=(\gamma \tilde{\pi}-m)
\Bigl[ \frac{\theta ((p_{0}+\mu)\sgn (p_{0}))}
{(\gamma\tilde{\pi})^2-m^2 + i \epsilon }+
\frac{\theta (-(p_{0}+\mu)\sgn (p_{0}))}
{(\gamma\tilde{\pi})^2-m^2 - i \epsilon }
\Bigr] .
\end{eqnarray}
Now, we  use the well known integral representation  of denominators
$$\frac{1}{\alpha \pm i0}=\mp i\int_{0}^{\infty}ds\e^{\pm i\alpha s},$$
which corresponds to introducing the ''proper--time'' $s$ into the
calculation
of the Eulier--Hei\-sen\-berg lagrangian by the Schwinger method \cite{schwin}.
We obtain
\begin{eqnarray}
\hat{G}=(\gamma \tilde{\pi}-m)
\Bigl[ &-&i \int_{0}^{\infty} ds \exp\left( i s \left[
(\gamma\tilde{\pi})^2-m^2 + i \epsilon \right]\right)
\theta ((p_{0}+\mu)\sgn (p_{0}))+\nonumber\\&+&
i \int_{0}^{\infty} ds \exp\left( -i s \left[
(\gamma\tilde{\pi})^2-m^2 - i \epsilon \right]\right)
\theta (-(p_{0}+\mu)\sgn (p_{0}))
\Bigr] .
\end{eqnarray}
For simplicity, we restrict ourselves only to the magnetic field case,
where $A_{0}=0, [\tilde\pi_{0},\tilde\pi_{\mu}]=0 $. Then we easily
can factorize the time dependent part of Green function
\be
G(x,x^{'})=\int \frac{d^3 p}{(2\pi)^3}\hat G \e^{ip(x-x^{'})}
=\int \frac{d^2 p}{(2\pi)^2}\hat G_{\vec{x}}
\e^{i\vec{p}(\vec{x}-\vec{x}^{'})}
\int \frac{d p_0}{2\pi}\hat G_{x_0} \e^{ip_0 (x_0 -x_0^{'})}.
\ee
By using the obvious relation
\begin{equation}
\label{yr}
(\gamma\tilde{\pi})^2=(p_0 +\mu)^2 -\vec{\pi}^2 +\frac{1}{2}g
\sigma_{\mu\nu}F^{\mu\nu}
\end{equation}
one gets
\begin{eqnarray}
&\*&G(x,x^{'})|_{x\rightarrow x^{'}}= -i\int\frac{dp_{0}}{2\pi}
\frac{d^2 p}{(2\pi)^2}(\gamma \tilde{\pi}-m)\int_{0}^{\infty} ds
\Biggl[
\e^{is(\tilde{p}_{0}^{2}-m^2)}\e^{-is\vec{\pi}^2}
\e^{isg\sigma F/2}-\nonumber\\
&-&
\theta (-(p_{0}+\mu)\sgn (p_{0}))\left(
\e^{is(\tilde{p}_{0}^{2}-m^2)}\e^{-is\vec{\pi}^2}
\e^{isg\sigma F/2}  +
\e^{-is(\tilde{p}_{0}^{2}-m^2)}\e^{is\vec{\pi}^2}
\e^{-isg\sigma F/2}\right)
\Biggr] .
\end{eqnarray}
Here the first term corresponds to the usual $\mu$--independent case and
there are two additional $\mu$--dependent terms.
In the calculation of the current the following trace arises:
\be
\tr \left[ \gamma^{\mu} (\gamma\tilde{\pi}-m)
\e^{isg\sigma F/2} \right]=2\pi^{\nu}g^{\nu\mu}\cos \left( g|^{*}\! F|s\right)
+2\frac{\pi^{\nu}F^{\nu\mu}}{|^{*}\! F|}\sin \left( g|^{*}\! F|s\right) -
2im\frac{^{*}\! F^{\mu}}{|^{*}\! F|}\sin \left( g|^{*}\! F|s\right),\nonumber
\ee
where $\*^{*}\! F^{\mu}=\varepsilon^{\mu\alpha\beta}F_{\alpha\beta}/2$
and $ |^{*}\! F|=\sqrt{B^2-E^2} $.
Since we are interested in  calculation of the
parity odd part (CS term)
it is enough to consider only
terms proportional to the dual strength tensor $\*^{*}\! F^{\mu}$.
On the other hand the term $2\pi^{\nu}g^{\nu\mu}\cos \left( g|^{*}\! F|s\right)$
at $\nu=0$ (see expression for the trace,
we take in mind that here there are only magnetic field)
also gives nonzero contribution to
the current $J^{0}_{c.s.}$ \cite{tolpa}
\be
J^{0}_{\rm even}=\frac{|eB|}{2\pi}\left( {\rm Int}\left[
\frac{\mu^{2}-m^{2}}{2|eB|}
\right]+\frac{1}{2}\right) \theta (|\mu | -|m|).
\ee
This part of current is parity invariant because under parity
$B\rightarrow -B$.
It is clear that this parity even object
does  contribute
neither to the parity anomaly nor to the mass of the
gauge field.
Moreover,  this term has been obtained \cite{tolpa}  in the pure
magnetic background and scalar magnetic field occurs
in the argument's denominator of the cumbersome function --
integer part. So, the parity even term seems to be
''noncovariantizable'', i.e. it can't be converted in covariant
form in effective action.
For a pity,  in  papers \cite{tolpa} charge density
consisting of both parity odd and parity even parts
is dubbed CS, what leads to misunderstanding.
The main goal of this paper is to explore the parity anomalous
topological CS term in the effective action at finite density. So, just the
term proportional to the dual strength tensor $\*^{*}\! F^{\mu}$ will
be considered.
The relevant part of the current reads
\begin{eqnarray}
J^{\mu}_{CS}&=&\frac{g}{2\pi} \int dp_{0}\int \frac{d^2 p}{(2\pi)^2}
\int_{0}^{\infty} ds\*
\frac{2im^{*}\! F^{\mu}}{|^{*}\! F|}\sin \left( g|^{*}\! F|s\right)
\Biggl[
\e^{is(\tilde{p}_{0}^{2}-m^2)}\e^{-is\vec{\pi}^2}
-\nonumber\\
&-&
\theta (-(p_{0}+\mu)\sgn (p_{0}))\left(
\e^{is(\tilde{p}_{0}^{2}-m^2)}\e^{-is\vec{\pi}^2} -
\e^{-is(\tilde{p}_{0}^{2}-m^2)}\e^{is\vec{\pi}^2}
\right)\Biggr] .
\end{eqnarray}
Evaluating two--momentum integral we derive
\begin{eqnarray}
J^{\mu}_{CS}=\frac{g^2}{4\pi^2} m \*^{*}\! F^{\mu}
\int_{-\infty}^{+\infty}\!\! dp_{0} \int_{0}^{\infty}\!\! ds
\Biggl[
\e^{is(\tilde{p}_{0}^{2}-m^2)}-
\theta (-\tilde{p}_{0}\sgn (p_{0}))\left(
\e^{is(\tilde{p}_{0}^{2}-m^2)}+
\e^{-is(\tilde{p}_{0}^{2}-m^2)m}\right)\Biggr] .
\end{eqnarray}
Thus, we get besides the usual CS part \cite{redl}, also
the $\mu$--dependent one.
It is easy to calculate it by use of the formula
$$ \int_{0}^{\infty}ds \e^{is(x^2-m^2)}=\pi\left(
\delta (x^2-m^2) +\frac{i}{\pi} {\cal{P}} \frac{1}{x^2-m^2}\right) $$
and we  get eventually
\begin{eqnarray}
\label{fab}
J^{\mu}_{CS}&=&\frac{m}{|m|}\frac{g^2}{4\pi}\*^{*}\! F^{\mu}
\bigl[ 1 - \theta (-(m+\mu)\sgn(m))
-\theta (-(m-\mu)\sgn(m))\bigr]\nonumber\\
&=&\frac{m}{|m|}\theta (m^2 -\mu^2 )\frac{g^2}{4\pi}\*^{*}\! F^{\mu}.
\end{eqnarray}

Let's now discuss the non-abelian case. Then  $A^{\mu}=T_{a} A_{a}^{\mu}$
in (2) and
$$\langle J_{a}^{\mu}\rangle= -i g \tr\left[\gamma^{\mu} T_{a}
G(x,x^{'})\right]_{x\rightarrow x^{'}} .$$
It is well--known \cite{redl,red11}
that there exist only two types of the constant background fields.
The first is the  ''abelian'' type
(it is easy to see that the self--interaction
$f^{abc}A_{b}^{\mu}A_{c}^{\mu}$
disappears under that choice of the background field)
\begin{equation}
\label{ab}
A_{a}^{\mu}=\eta_{a}\frac{1}{2}x_{\nu}F^{\nu\mu},
\end{equation}
where $\eta_{a}$ is an arbitrary constant
vector in the color space, $F^{\nu\mu}={\rm Const}$.
The second  is the pure ''non--abelian'' type
\begin{equation}
\label{nab}
A^{\mu}={\rm Const}.
\end{equation}
Here the derivative terms (abelian part) vanish from the strength tensor
and it contains only the self--interaction part
$F^{\mu\nu}_{a}=gf^{abc}A_{b}^{\mu}A_{c}^{\mu}$.
It is clear that to catch  abelian part of the CS term
we should consider the background field (\ref{ab}),
whereas for the non--abelian (derivative noncontaining,
cubic in $A$) part
we have to use the case (\ref{nab}).

Calculations in the ''abelian'' case reduces to the previous analysis,
except the trivial adding  of the color indices in the formula (\ref{fab}):
\begin{eqnarray}
\label{finab}
J^{\mu}_{a}
=\frac{m}{|m|}\theta (m^2 -\mu^2 )
\frac{g^2}{4\pi}\*^{*}F^{\mu}_{a} .
\end{eqnarray}
In the case (\ref{nab})  all calculations are similar. The only
difference is that the origin of term $\sigma_{\mu\nu}F^{\mu\nu}$
in (\ref{yr}) is not the linearity $A$ in $x$ (as in abelian case) but
the pure non--abelian   $A^{\mu}={\rm Const}$. Here  term
$\sigma_{\mu\nu}F^{\mu\nu}$ in (\ref{yr}) becomes quadratic in $A$
and we have
\begin{eqnarray}
\label{finnab}
J^{\mu}_{a}
=\frac{m}{|m|}\theta (m^2 -\mu^2 )
\frac{g^3}{4\pi}\varepsilon^{\mu\alpha\beta}
\tr\left[ T_{a}A^{\alpha}A^{\beta} \right] .
\end{eqnarray}
Combining formulas (\ref{finab}) and (\ref{finnab}) and integrating
over field $A^{\mu}_{a}$  we obtain eventually
\begin{equation}
S^{{\rm C.S}}_{eff}=\frac{m}{|m|}\theta (m^2 -\mu^2 ) \pi W[A] ,
\end{equation}
where $W[A]$ is the CS term
$$W[A]=\frac{g^2}{8\pi^2}\int d^{3}x \varepsilon^{\mu\nu\alpha}
\tr \left( F_{\mu\nu}A_{\alpha}-
\frac{2}{3}gA_{\mu}A_{\nu}A_{\alpha}\right) .$$

This result can be obtained also with an arbitrary initial field
configuration by use of the perturbative expansion.
Here we work at once in the nonabelian case.
The effective action looks as
\begin{eqnarray}
S_{eff} &=&
\frac{1}{2}\tr\int_{x}  A_{\mu}(x)\int_{p}\e^{-ixp}A_{\nu}(p)
\Pi^{\mu\nu}(p)\nonumber\\
&+&\frac{1}{3}\tr\int_{x}  A_{\mu}(x)\int_{p,r}\e^{-ix(p+r)}
A_{\nu}(p)A_{\alpha}(r)\Pi^{\mu\nu\alpha}(p,r),
\end{eqnarray}
where polarization operator and vertice have the standard form
\begin{eqnarray}
\Pi^{\mu\nu}(p) &=&g^2 \int_{k} \tr \left[ \gamma^{\mu}
G(p+k;\mu)\gamma^{\nu}G(k;\mu)\right] \nonumber\\
\Pi^{\mu\nu\alpha}(p,r) &=& g^3
\int_{k}\tr \bigl[ \gamma^{\mu}G(p+r+k;\mu)
\gamma^{\nu}G(r+k;\mu)\gamma^{\alpha}G(k;\mu)
\bigr] .
\end{eqnarray}
First consider the second order term.
Signal for the mass generation (CS term) is
$\Pi^{\mu\nu}(0)\not =0$.
Picking up a parity odd part  we get
\begin{eqnarray}
\Pi^{\mu\nu}=g^2 \int_{k} ( -i2m e^{\mu\nu\alpha} p_{\alpha} )
\frac{1}{(\tilde{k}^2 -m^2+i \epsilon (k_{0}+\mu)\sgn (k_{0}))^2}.
\end{eqnarray}
After some simple algebra one obtains
\begin{eqnarray}
\Pi^{\mu\nu}=-i\frac{m}{|m|}\theta (m^2 -\mu^2 )\frac{g^2}{4\pi}e^{\mu\nu\alpha} p_{\alpha}.
\end{eqnarray}
In the same manner, handling the third order contribution one gets
\begin{eqnarray}
\Pi^{\mu\nu\alpha}
=-i\frac{m}{|m|}\theta (m^2 -\mu^2 )\frac{g^3}{4\pi}e^{\mu\nu\alpha}.
\end{eqnarray}
Substituting  vertices in the effective action we
eventually get
\begin{eqnarray}
S^{C.S.}_{eff} =\frac{m}{|m|}
\theta (m^2 -\mu^2 ) \frac{g^2}{8\pi}
\int d^{3}x e^{\mu\nu\alpha}
\tr\left(
A_{\mu}\partial_{\nu}A_{\alpha}-
\frac{2}{3}g A_{\mu}A_{\nu}A_{\alpha}\right)  .
\end{eqnarray}
So, we obtain the same CS $\mu$--dependent  coefficient
as in the previous method at once in the effective action.

Moreover, by use of the perturbative theory we also derive
CS term at finite density in 5-dimensional nonabelian
gauge theory. All calculations are similar to 3--dimensional
case (details will be published elsewhere)
\be
S_{eff} &=&\frac{m}{|m|}
\theta (m^2-\mu^2 ) \frac{g^3}{48\pi^2}
\int d^{5}x e^{\mu\nu\alpha\beta\gamma}\nonumber\\&\times& \tr\Bigl(
A_{\mu}\partial_{\nu}A_{\alpha}\partial_{\beta}A_{\gamma}
+\frac{3}{2}g A_{\mu}A_{\nu}A_{\alpha}\partial_{\beta}A_{\gamma}+
\frac{3}{5}g^{2} A_{\mu}A_{\nu}A_{\alpha}A_{\beta}A_{\gamma}
\Bigr).
\ee
From the above direct calculations  it is clearly seen
that the chemical potential dependent coefficient is the
same for all parity odd parts of action and doesn't depend on
space dimension:
\begin{equation}
\label{kon}
S^{{\rm C.S}}_{eff}=\frac{m}{|m|}\theta (m^2 -\mu^2 ) \pi W[A] ,
\end{equation}
where $W[A]$ is the CS term in any odd dimension.
Since  only the lowest orders
of perturbative series contribute to CS term at finite density
(the same situation
is well-known at zero density), the result obtained by using
formally perturbative technique appears to be nonperturbative.
Thus, the $\mu$--dependent CS term coefficient
reveals the amazing property of universality.
Namely, it does depend on
neither dimension of the theory nor abelian or nonabelian gauge
theory is studied.

%%%%%%%%%%%%%%%%%%%%%%%%%%%%%%%%%%%%%%%%%%%%%%%%%%%%%%%%%%%%%%%%%%%%%%%%%%%%%
The arbitrariness of $\mu$ gives us the possibility
to see Chern--Simons coefficient behavior at any masses.
It is very interesting that  $\mu^2 = m^2$ is the
crucial point for Chern--Simons.
Indeed, it is clearly seen from (\ref{kon}) that when $\mu^2 < m^2$
$\mu$--influence disappears and we get the usual Chern-Simons term
$I^{{\rm C.S}}_{eff}= \pi W[A].$
On the other hand when $\mu^2 > m^2$
the situation is absolutely different.
One can see that here the Chern-Simons term
disappears because of non--zero density of background fermions.
We'd like to emphasize the
important massless case $m=0$ considered in \cite{ni1}.
Then even negligible density,
which always take place in any
physical processes, leads to vanishing of the parity anomaly.

In conclusion, let us stress again that we nowhere have used
any restrictions on $\mu$.
Thus we not only confirm result
in \cite{ni1} for Chern--Simons in $QED_{3}$ at small density,
but also expand it
on arbitrary $\mu$, nonabelian case and arbitrary odd dimension.

%%%%%%%%%%%%%%%%%%%%%%%%%%%%%%%%%%%%%%%%%%%%%%%%%%%%%%%%%%%%%%%%%%%%%%%%%%%%%

We would like to thank V.Fedyanin,  V.Osipov and O.Veretin
for helpful discussions.

\end{document}